\documentstyle[12pt,epsfig,aps]{revtex}

\newcommand{\bce}{\begin{center}}
\newcommand{\ece}{\end{center}}
\newcommand{\beq}{\begin{equation}}
\newcommand{\eeq}{\end{equation}}
\newcommand{\bea}{\begin{eqnarray}}
\newcommand{\eea}{\end{eqnarray}}
\newcommand{\ba}{\begin{array}}
\newcommand{\ea}{\end{array}}

\newcommand{\doublespace}{
    \renewcommand{\baselinestretch}{1.6}\large\normalsize}

\def\lsim{\mathrel{\rlap{\lower4pt\hbox{\hskip1pt$\sim$}}
    \raise1pt\hbox{$<$}}}         
\def\gsim{\mathrel{\rlap{\lower4pt\hbox{\hskip1pt$\sim$}}
    \raise1pt\hbox{$>$}}}         

\def\Pom{{\bf I\!P}}

\def\PLB{{Phys. Lett.}  B}

\def\ZPC{{Z. Phys.} C}

 \advance\oddsidemargin   by -1in
 \advance\evensidemargin  by -1in
 \advance\topmargin   by -0.5in
\begin{document}
\vspace{.4cm}

\begin{center}
{\Large \bf Charged Current Diffractive Structure Functions
\vspace{1.0cm}\\}
{\large \bf  \underline {M. Genovese}\footnote{ \small
Supported by the EU Program ERBFMBICT 950427} $^{a}$, 
M. Bertini$^{b}$, N.N.~Nikolaev$^{c}$,
B.G.~Zakharov$^{c}$ \vspace{1.0cm}\\}
{\it

$^{a}$  Institut des Sciences Nucl\'eaires \\
Universit\'e Joseph Fourier--IN2P3-CNRS, \\
53, avenue des Martyrs, F-38026 Grenoble Cedex,
France
\bigskip \\
$^{b}$ INFN, Sezione di Torino, Via P.Giuria 1, I-10125 Torino, Italy
\bigskip \\
$^{c}$IKP(Theorie), KFA J{\"u}lich, 5170 J{\"u}lich, Germany
\medskip\\
and L. D. Landau Institute for Theoretical Physics, GSP-1,
117940, \\
ul. Kosygina 2, Moscow V-334, Russia.\vspace{1.0cm}\\ }

\vspace{1cm}

{\Large \bf Abstract} \\
\end{center}
We present our study of the diffraction in charged current DIS. 
We analyse the perturbatively
tractable excitation of heavy quarks,  emphasizing the peculiarities
of the Regge factorization breaking in excitation of open charm.
\bigskip

\bigskip

\begin{center}
E-mail: genovese@isnhp4.in2p3.fr
\end{center}

\vspace{2cm}

\doublespace


The study of diffractive electromagnetic Deep
Inelastic Scattering in QCD has rapidly advanced in last years
(\cite{NZ92,Levin},
for a recent review see \cite{NZDIS97}). Successful quantitative
predictions for diffractive DIS are now available \cite{GNZ95,HT4,NZDIS97},
the pQCD factorization scales for different components of the diffractive
Structure Function 
have been derived \cite{GNZcharm,GNZlong,Bartels},
breaking of the diffractive factorization and of the DGLAP evolution for
diffractive SF's have been understood and a large perturbative
intrinsic charm in the pomeron
has been established \cite{GNZ95,GNZcharm,GNZlong}, 
the prediction of large perturbative
higher twist contributions has been obtained \cite{GNZlong,HT4}. 

The study of diffraction in charged current (CC) DIS, $e p\rightarrow
\nu p'X$, will permit a further test of this approach.
Rapidity gap events in CC DIS have already been observed
at HERA \cite{CCZEUS}
and with amassing more data on CC DIS a detailed comparison
between the experiment and models for diffractive DIS will be
possible \cite{CCDDexp}. Because of the parity non-conservation, in the CC case
one has a larger variety of diffractive SF's
compared to the neutral current  electromagnetic (EM) case.
To the lowest order in pQCD, CC diffractive DIS proceeds by the
Cabibbo-favoured excitation of the $(u\bar{d})$ and $c\bar{s}$
dijet states. The unequal mass $(c\bar{s})$ is a particularly
interesting testing ground where to investigate diffractive factorization
breaking. A self-tagging property of charm
jets gives better access to various diffractive structure
functions, for instance, to $F_{3}^{D(3)}$.

The subject of this proceeding is the derivation of the above stated
features of CC diffractive DIS and the discussion of its differences 
respect to EM
diffractive DIS. For convenience we focus on the process  
$e^{+}p\rightarrow \bar{\nu} p'X$ already observed by ZEUS \cite{CCZEUS}. 
The discussion and results can easily be translated to the 
$e^{-} p\rightarrow {\nu} p'X$ process.

We start with necessary definitions. In diffractive CC $e^{+}p$ 
scattering the experimentally 
measured quantity is the five-fold differential cross section
$d\sigma^{(5)}(ep\rightarrow \nu p'X)/
dQ^{2}dxdM^{2}dp_{\perp}^{2}d\phi$. Here $X$ is the diffractive
state of mass $M$, $p'$ is the secondary proton with the transverse
momentum $\vec{p}_{\perp}$, $t=-\vec{p}_{\perp}^{2}$, $\phi$ is the 
angle between the $(e,e')$ and $(p,p')$ planes, $Q^{2}=-q^2$ is the 
virtuality of the $W^{+}$ boson, $x, y, x_{\Pom}$ and 
$\beta=x/x_{\Pom}$ are the standard diffractive DIS variables.

The underlying subprocess is diffraction excitation of the $W^+$
boson, $W^+ p\rightarrow p'X$. In the parity conserving EM DIS, the
exchanged photon have either longitudinal (scalar), 
$s={1 \over Q} (q_{+}n_{+} - q_{-} n_{-})$ or
transverse, in the $(e,e')$ plane, polarization $t_{\mu}$ (here
$n_{\pm}$ are the usual light--cone vectors, $n_{+}^{2}=n_{-}^{2}=0$,
$n_{+}n_{-}=1, q=q_{+}n_{+}+q_{-}n_{-}$ and $q.s = q.t = 0$). In the
parity-nonconserving CC DIS, the exchanged $W^{+}$ bosons have also
the out-of-plane linear polarization $w_{\mu}=\epsilon_{\mu\nu\rho\sigma}
t_{\nu}n_{\rho}^{+}n_{\sigma}^{-}$. We introduce also the usual
transverse metric tensor 
$\delta_{\mu \nu}^{\perp} =\delta_{\mu\nu}
+n_{\mu}^{-}n_{\nu}^{+} + n_{\mu}^{+}n_{\nu}^{-}= - t_{\mu} t_{\nu}
-w_{\mu} w_{\nu}$. Then, the polarization state of the $W^{+}$
is described by the leptonic tensor

\bea
{L_{\mu \nu} \over 4}= {2 Q^2 \over  y^2} \left[ -{1\over 2}
\delta^{\perp}_{\mu\nu}(1-y+{1\over 2}y^2)
+ {1 \over 2}(1-y) (t_\mu t_\nu - w_\mu w_\nu) +
(1-y) s_\mu s_\nu \;\; \right . \nonumber\\ 
\left . + {1 \over 2} (1-{1\over 2} y) \sqrt{ 1- y}(s_\mu t_\nu+ s_\nu t_\mu ) +
 {i \over 2} y (1- {1 \over 2} y) (w_\mu t_\nu - w_\nu t_\mu) +
 {i \over 2} y \sqrt{1-y} (w_\mu s_\nu - s_\mu w_\nu) \right ]
\label{eq:Lep}
\eea
which, upon contraction with the hadronic tensor leads to
6 different components for $d\sigma_{i}^{(3)}
(W^{+}p\rightarrow p'X)/dM^{2}dt d\phi$
($i=T,L,TT',LT,3$ and $LT(3)$):

\bea
{y d\sigma^{(5)}(e^{+}p\rightarrow \bar{\nu} p'X)\over
dQ^{2}dydM^{2}dp_{\perp}^{2}d\phi} =
{G_{F} M_W^{2} Q^{2}  \over 4 \sqrt{2}\pi^2 (M_{W}^{2}+Q^2)^2}
\left\{(1-y+{1\over 2}y^{2})\cdot d\sigma_{T}^{D(3)}
-y(1-{1\over 2}y)\cdot d\sigma_{3}^{D(3)}\right. \nonumber\\
+(1-y)\cdot d\sigma_{L}^{D(3)}+
(1-y)\cos 2\phi \cdot d\sigma_{TT'}^{D(3)}\nonumber\\
\left.
+(1-{1\over 2}y)\sqrt{1-y}\cos\phi
\cdot d\sigma_{LT}^{D(3)}
 -y\sqrt{1-y}\sin\phi \cdot d\sigma_{LT{(3)}}^{D(3)}\right\}
/dM^{2}dp_{\perp}^{2}d\phi\, ,
\label{eq:SigDIS}
\eea
where $G_{F}$ is the Fermi coupling, $m_{W}$ is the mass of the 
$W$-boson.
One can then define the dimensionless diffractive structure functions $F_{i}^{D(4)}$,

\beq
{(Q^{2}+M^{2})d\sigma_{i}^{(3)}(W^{+}p\rightarrow p'X)\over
dM^{2}dp_{\perp}^{2}}= 
{\pi G_{F} M_{W}^{2} Q^2 \over \sqrt{2}(Q^{2}+M_W^2)^2}\cdot 
{\sigma_{tot}^{pp}\over 16\pi}\cdot
F_{i}^{D(4)}(p_{\perp}^{2},
x_{\Pom},\beta,Q^{2})\, ,
\label{eq:Fdiff4}
\eeq
It is also useful to introduce the  t--integrated SF's 
\footnote{The ZEUS/H1 definition \cite{ZEUSF2Pom} corresponds to
$F_{2}^{D(3)}(H1/ZEUS)={1\over x_{\Pom}}F_{2}^{D(3)}$, which blows up
at $x_{\Pom}\rightarrow 0$, and is much less convenient than our
$F_{2}^{D(3)}$.}

\beq
F_{i}^{D(3)}(x_{\Pom},\beta,Q^{2})=
{\sigma_{tot}^{pp}\over 16\pi }\int  dp_{\perp}^{2}
F_{i}^{D(4)}(p_{\perp}^{2},x_{\Pom},\beta,Q^{2})\, .
\label{eq:Fdiff3}
\eeq
The diffractive SF's $F_{T}^{D(3)}, F_{L}^{D(3)}$ and
$F_{3}^{D(3)}$ are counterparts of the familar $F_{T}=F_{2}-F_{L},
F_{L}$ and $F_{3}$ for inclusive DIS of neutrinos,
$F_{3}^{D(3)}$ and $F_{LT(3)}^{D(3)}$, are C- and P-odd and vanish
in EM scattering. In the following we will not consider the azimuthal
angle-dependent terms, but will limit our discussion to 
$F_{T}^{D(3)}, F_{L}^{D(3)}$ and $F_{3}^{D(3)}$.

Only relatively large $x \sim 10^{-2}$ are easily accessible in CC
DIS \cite{CCZEUS,CCDDexp}. As the experimental selection of diffractive
events requires $x_{\Pom} <$0.05-0.1, the kinematical relation
$\beta=x/x_{\Pom}$ implies that the experimentally observed CC
diffractive DIS will proceed at rather large $\beta$,
dominated by the partonic subprocess $W^{+}p\rightarrow (u\bar{d}) p',
(c\bar{s}) p'$. 

We focus on the $c\bar{s}$ excitation, analogous considerations 
apply to $u \bar{d}$. In what follows, $z$ and $(1-z)$ are the fractions 
of the (light--cone) 
momentum of the $W^{+}$ carried by the charmed quark and strange
antiquark respectively, $\vec{k}$ is the relative transverse momentum in
the $q\bar{q}$ pair. The invariant mass of the di-jet final states equals

\beq
M^2={\frac{k^2 + \mu^2}{z(1-z)}}  \, ,
\label{eq:M2}
\eeq
where $\mu^2=(1-z)m_c^2 + zm_s^2$. $m_c, (m_s)$ being the 
charmed (strange) quark mass. 
All SF's are calculable in terms of the same  quark helicity
changing and conserving amplitudes $\vec{\Phi}_{1}$ and $\Phi_{2}$
introduced in \cite{NZ92,NZsplit}. Combining the formalism of \cite{NZsplit}
with the treatment of charm leptoproduction in \cite{F2cs}, we 
obtain (integration over the azimuthal orientation of $\vec{k}$ is
understood, $\alpha_{cc}={G_F M^2_W \over 2 \pi \sqrt{2}}$)

\bea
\frac{d\sigma^D_{L,T}}{dz dk^2 dt}\bigg |_{t=0} =
\frac{\pi^2 \alpha_{cc}}{3} 
\alpha_{S}^{2}(\bar{Q}^{2})
\left[ 
A_{L,T}\, (z,m_s,m_c)\vec{\Phi}_{1}^{2}  +
B_{L,T}\, (z,m_s,m_c)\Phi_{2}^{2}
\right] \, ,
\label{eq:DSigLT}
\eea
(see later for a definition of $\bar{Q}^{2}$), where
\bea
A_T(z) &=& \left[ 1 -2z(1-z) \right] \, ,
\label{eq:AT}
\\
B_T(z,m_{s},m_{c}) &=& \left[ m_{c}^2 - 2 z(1-z) m_{c}^2 - z^2 \Delta^2 \right]\, ,
\label{eq:BT}
\\
A_3(z) &=& \left[2z - 1  \right] \, ,
\label{eq:A3}
\\
B_3(z,m_{s},m_{c}) &=& \left[z^2 m_{s}^2- (1-z)^{2}m_{c}^{2}\right]\, ,
\label{eq:B3}
\\
A_L(z,m_{s},m_{c}) &=&{1\over Q^{2}}(m_{s}^{2}+m_{c}^{2}) \, ,
\label{eq:AL}
\\
B_L(z,m_{s},m_{c}) &=& 4Q^2 z^2 (1-z)^2
+4z(1-z)\mu^2 +{1\over  Q^{2}}[\mu^4+m_{c}^{2}m_{s}^{2}]
\label{eq:BL}
\eea
with $\Delta^2 = m_c^2 -m_s^2$. 

The amplitudes $\vec{\Phi}_{1}$ and $\Phi_{2}$ were
derived in \cite{NZ92,NZsplit}:
\bea
\vec{\Phi}_{1} =\vec{k} \int \frac{d\kappa^{2}}{\kappa^{4}}
f(x_{\Pom},\kappa^{2})
\left[{1\over k^{2}+\varepsilon^{2}}-{1\over \sqrt{a^{2}-b^{2}}}+
{2\kappa^{2} \over a^{2}-b^{2}+a\sqrt{a^{2}-b^{2}}}\right]\nonumber\\
\approx {2\vec{k} \varepsilon^{2} \over (k^{2}+\varepsilon^{2})^{3}}
\int {d\tau \over \tau}W_{1}(\omega,\tau)G(x_{\Pom},\tau \bar{Q}^{2})
\, ,
\label{eq:phi1}
\\
\Phi_{2}=\int \frac{d\kappa^{2}}{\kappa^{4}}
f(x_{\Pom},\kappa^{2})
\left[{1\over \sqrt{a^{2}-b^{2}}} -
{1\over k^{2}+\varepsilon^{2}}\right]\nonumber\\
 \approx {k^{2}-\varepsilon^{2} \over (k^{2}+\varepsilon^{2})^{3}}
\int {d\tau \over \tau}W_{2}(\omega,\tau)G(x_{\Pom},\tau \bar{Q}^{2})
\, ,
\label{eq:phi2}
\eea
Here
$f(x,\kappa^{2})=\partial G(x,k^{2})/\partial \log(\kappa^{2})$
($G(x,Q^2)=x g(x,Q^2)$)
is the unintegrated gluon SF of the proton,
$\varepsilon^{2}=z (1-z) Q^2 + zm_s^2 + (1-z) m_c^2,a=\varepsilon^{2}+
k^{2}+\kappa^{2}, b=2k\kappa$ and

\beq
\bar{Q}^{2} = \varepsilon^{2}+k^{2} ={ k^{2} + zm_{s}^{2} + (1-z)m_{c}^{2} \over 1-\beta }\,,
\label{eq:Q2scale}
\eeq
emerges as the pQCD factorization scale. 
A second representation, in which
\bea
W_1(\omega, \tau) &=& \frac{(1 + \omega^2)}{2\tau(1+\omega)}
\left \{
\left [ 1- \frac{\Theta
+\tau^2[1+(1-\omega)/\tau(1+\omega)]}{(1+w)\Theta^{3/2}}
\right ]  \right. \hspace{3cm} \nonumber \\
\nonumber \\
&& + \left.
\frac{4\tau^2(1+\tau)}{\Theta [\Theta^{1/2} + (1+\tau)]^2}
\left [ 1 + \frac{1+\tau (1+\omega) -2\omega \tau/(1+\tau)}
{(1+\omega)\Theta^{1/2}}
\right] \right \}
\label{eq:W1}
\eea
\bea
W_2(\omega,\tau) &=&  \frac{(1 + \omega)}{\tau (1-\omega)}
\left[ 1 - \frac{\Theta +  \tau^2 [1 + (1-\omega)/\tau(1+\omega)}
{\Theta^{3/2}} \right]
\label{eq:W2}
\eea
where
\beq
\Theta= (1+\omega)^2 \left[ (1+ \tau) ^2-
{4 \omega \tau \over (1+\omega)} \right]\,.
\label{eq:Delta}
\eeq
$\tau=k^2/\bar{Q}^2$ and $\omega=k^2/\varepsilon^2$, is more convenient
for the practical calculations based on available parameterization
for the gluon SF of the proton $G(x,Q^{2})$. Here
the weight functions $W_{i}(\omega,\tau)$ have a narrow peak at
$\tau \approx 1 $ with the unit area under the peak, which gives
the Leading Log$Q^{2}$ result \cite{NZsplit,GNZcharm}
\beq
\int {d\tau \over \tau}W_{i}(\omega,\tau)G(x_{\Pom},\tau \bar{Q}^{2})
\approx G(x_{\Pom},\bar{Q}^{2})\, ,
\label{eq:Geff}
\eeq
valid for sufficiently large values of $\omega$, which
is equivalent to sufficiently large $\beta \gsim 0.1$ (of interest in 
the present study). For applications of the above formalism
to the EM case see \cite{GNZcharm,GNZlong,Bartels,Levin}. Also, despite
the somewhat different appearance, the so-called soft color
interaction model by Buchm\"uller et al. \cite{Buchmuller} is
essentially identical to the
above described picture of diffractive DIS (for more discussion on
that see \cite{NZDIS97}).

At variance with the equal mass EM case, where $\bar{Q}^{2}=(k^{2}+m^{2})/(1-\beta)$, 
now the factorization scale depends on $z$ and then one expects 
different cross sections whether the charmed
quark is produced in the forward (F) or the backward (B) hemisphere,
with respect to the $W$ momentum, in the rest frame of the diffractive
state $X$. The two configurations differ by the value of the light-cone
variable $z_{F,B}={1\over 2}(1+\delta)\left[1\pm
\sqrt{ 1 -4 {k^2 + m_c^2 \over M^2 (1 + \delta)^{2}}}\right]$,
where we have introduced for brevity the variable 
$\delta= {\Delta^2 \over Q^2}\; {\beta \over (1-\beta)}$.
The pQCD scale is  perturbatively large for large
$\beta$ even for light flavours, and for the charm component of the
diffractive SF it is large for all $\beta$ (see below).

\noindent 
Evaluating the light quark component of the diffractive SF 
at not so large $\beta$, one needs a model for the small-$Q^2$ 
behaviour of the 
gluon structure function $G(x,Q^{2})$: in the following we will use the
same form used in Ref. \cite{HT4}, which at large $k^2$ coincide with the 
GRV NLO parameterization \cite{GRVNLO}.

\noindent 
For what concerns quark masses we assume 
$m_c=1.5$ GeV, $m_s=0.3$ GeV and $m_{u,d}= 150$ MeV.
Variations of the charm mass
by 10$\%$ have a little effect on the predicted
SF, apart from shifting the threshold $\beta_{c}=
Q^{2}/[Q^{2}+(m_{c}+m_{s})^{2}]$ (see below).

For evaluating $F_{i}^{D(3)}$ one needs to know the
$p_{\perp}^{2}$ dependence of the diffractive cross section,
which is usually parameterized as $d\sigma/dp_{\perp}^{2}
\propto \exp(-B_{D}p_{\perp}^{2})$. As it was shown in
\cite{NPZslope,NZDIS97}, for heavy flavour excitation, for
the perturbative transverse higher twist and longitudinal contributions 
$B_{D}\sim 6 $GeV$^{-2}$, for light flavour contribution 
the diffraction slope $B_{D}$ exhibits, at not so large $\beta$, 
a slight $\beta$-dependence, but for the purposes of this present 
exploratory study we can simply 
take $B_D (ud)=9$ GeV$^{-2}$.

Various studies of diffractive DIS (as DIS off pomerons
in the proton) have  assumed diffractive
factorization. The latter is not supported by QCD studies
\cite{GNZ95,GNZcharm}, and the present study of charm excitation
in CC diffractive DIS offers more
evidence to this effect. Nevertheless, wherever that would not lead to a grave
confusion, we will speak of the perturbative intrinsic partons in the pomeron.

Separation of the pQCD subprocess of $W^{+}\rightarrow c\bar{s}$ into the
excitation of charm on the perturbative intrinsic strangeness in the pomeron and
excitation of (anti)strangeness on the intrinsic (anti)charm is not
unambiguous and must be taken with the grain of salt. In the naive
parton model, in the former process charmed quark will carry the
whole momentum of the $W^{+}$ and be produced with $z\approx1$.
In contrast, in the latter process, it is the strange antiquark
which carries the whole momentum of $W^{+}$ and charmed quark is
produced with $z\approx 0$, which suggests $z>{1\over 2}$ and
$z<{1\over 2}$ as a compromise boundary between the two partonic
subprocesses. However, the full fledged pQCD calculation
leads to broad $z$ distributions (for a related discussion of definition
of the strangeness and charm density in $\nu N,\bar{\nu}N$ inclusive DIS
see \cite{F2cs}). As a purely operational definition, we stick to a
parton model decomposition $F_{T}^{D(3)}=F_{T(s)}^{D(3)}+
F_{T(\bar{c})}^{D(3)}$ and $F_{3}^{D(3)}=F_{T(s)}^{D(3)}-
F_{T(\bar{c})}^{D(3)}$, which is a basis for the results shown in
Fig.~1. 
\begin{figure}[htb] \label{FIGUR1}
\mbox{\epsfig{file=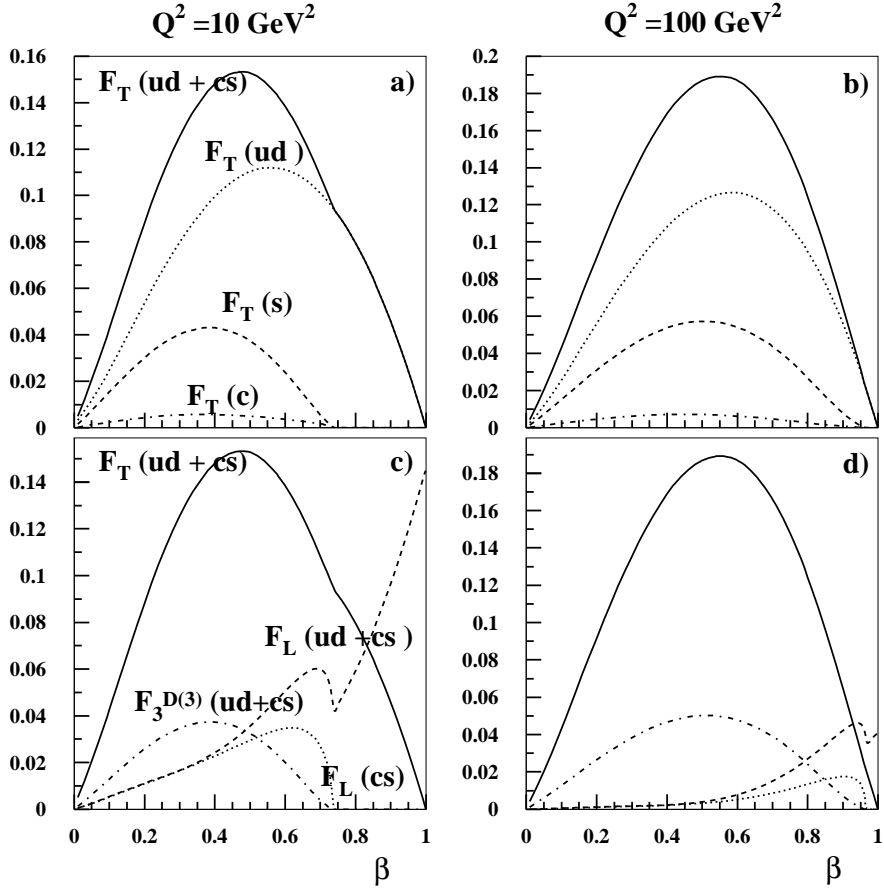,width=0.8\textwidth}}
\caption{The $\beta$ dependence for $x_{\Pom}=10^{-3}$:
a) $F_T (u,d,s,c)$ [solid], $F_T (u,d)$ [dotted], $ F_T(s)$
[dashed], $F_T(c)$ [dot--dashed]   at $Q^2 =  10$GeV$^2$
b)  the same as above for  $Q^2 =100$GeV$^2$
c) All flavours    $ F_T$ [solid], $F_3$ [dot--dashed], $F_L$
[dashed] and $F_L(cs)$ [dotted]    at  $Q^2 = 10$GeV$^2$
d)  the same as above for  $Q^2 = 100$GeV$^2$
} 
\end{figure}

With this definition, excitation of the
charmed quark off the intrinsic strangeness, $F_{T(s)}^{D(3)}$,
comes from terms $\propto z^2$ in (\ref{eq:AT}),(\ref{eq:BT})
and (\ref{eq:A3}),(\ref{eq:B3}). It is dominated by the forward
production of charm w.r.t. the momentum of $W^{+}$ in the
rest frame of the diffractive system $X$, but receives certain
contribution also from $z< {1\over 2}$. Similarly, $F_{T(\bar{c})}^{D(3)}$
come from terms $\propto (1-z)^2$, is dominated by the forward
production of strangeness (the backward production of charm), but
still receives certain contribution from the forward charm production.

\bigskip

All the considerations of Ref. \cite{GNZlong,HT4} for the longitudinal 
and transverse diffractive SF in electroproduction are fully
applicable to the CC case at $Q^{2}\gg m_{c}^{2}$. 
We consider first the backward charm, $z\ll 1$, for which Eq.s 
(\ref{eq:M2}, 
\ref{eq:Q2scale}) give  $z \approx (k^2 + \mu^2)/M^2$ and 
$\bar{Q}^{2} \approx (k^{2}+m_{c}^{2})/(1-\beta)$.
Expanding the brackets of Eqs.(\ref{eq:phi1}, \ref{eq:phi2}), 
in Eq.(\ref{eq:DSigLT}) and using the approximation (\ref{eq:Geff}) 
the $k^2$-integration in (\ref{eq:DSigLT}) gives dominant contributions 
to the transverse SF coming from the low-$k^2$ region but without entering 
deeply in the nonperturbative region for the heavy quark production. For 
$M^2 \gg m_c^2$ one finds for the low scales dominated contribution 
(including the LT and the first HT):

\bea
F_{T(\bar{c}) }^{D(4)} &\approx&   \frac{4 \pi}{3\sigma^{pp}_{tot}}\, 
\frac{\beta (1-\beta)^{2}}{ 6 m_c^2 (1+\delta)} 
\left\{ (3 + 4\beta + 8\beta^2) + 
{m^2_c \over Q^2} \frac{4 \beta}{1-\beta}   \right. 
\\ \nonumber 
&.& \left.
 \times   \left[ {5 \over 4} {\Delta^2 \over m^2_c}(1+ 8\beta^2) 
- (1-2\beta+4\beta^2)\right] \right\}
\left[ \alpha_S (\bar{Q}^2_{L}) 
G(x_{\Pom},\bar{Q}_{L}^{2} \simeq {m_c^2 \over (1-\beta)}) \right]^2 
\label{eq:FTL}
\eea

As in the EM case, the large $k^2$  ($k^2 \sim M^2/4$) dominated 
contributions come from the second term in (\ref{eq:AT}). They are 
calculable in pQCD and the twist expansion starts with twist-4:  

\beq
F_{T(\bar{c}) }^{D(4)} = - \frac{16 \pi}{\sigma^{pp}_{tot}}\;
{2 \over 3} \frac{\beta^2 (1-\beta)}{Q^2 (1+\delta)} 
\left( \beta^2 + 2\; {\Delta^2 \over Q^2} \;  
\frac{\beta^2 (2\beta-1)}{(1-\beta)} \right)   
\left[\alpha_S (\bar{Q}^2_{H}) 
G(x_{\Pom},\bar{Q}_{H}^{2}\simeq {Q^2\over 4 \beta})\right]^2 
\label{eq:FTH}
\eeq

In (20,21) emerge additional higher-twist corrections 
$\propto (\Delta^2/Q^2)^n$, in a first approach we restrict ourselves to the 
genuine contributions and its first higher twist corrections.

The higher twist corrections to $F^D_T$ receive thus both 
 contributions from the 
low-scale region and the large scale $\bar{Q}^2_H$. The first term in 
Eq.(\ref{eq:FTH}) is substantially the same which has been discussed in 
Ref.\cite{HT4} for EM current and it remains relevant even at relatively large 
value of $Q^2$ as the $1/Q^2$ factor is partially compensated by the growth of 
gluon SF $G(x_{\Pom}, \bar{Q}^2_H)$.

Let us now consider the longitudinal cross section. The most important
contribution comes from the term $z^{2}(1-z)^{2}Q^{2}$ in the $B_{L}$
expansion (\ref{eq:BL}), which is identical to that
in the EM case. The $k^{2}$ integrated cross section is completely 
dominated by the short-distance contribution from high-$k^{2}$ jets,
$k^{2} \sim {1\over 4}M^{2}$. Upon the $k^{2}$ integration, to a
logarithmic accuracy, we find the twist expansion of the 
longitudinal SF for the terms dominated by the large scale:

\beq
F_{L}^{D(4)} = \frac{16 \pi}{\sigma^{pp}_{tot}}\; 
\frac{\beta^{3}\;(2\beta -1)}{3\ Q^{2}(1+\delta)}
\left((2\beta-1) + 
{\Delta^2\over Q^2}\frac{\beta(5 - 6\beta)}{(1-\beta)}\right)
\left[\alpha_S(\bar{Q}^2_H) G(x_{\Pom},\bar{Q}^{2}_H ) \right]^2
\label{eq:FLH}
\eeq

As the pQCD factorization scale, $\bar{Q}_{H}^{2}$, does not depend on 
flavours we predict a restoration of the flavours symmetry 
at asymptotically large $Q^{2}$.
Again the scaling 
violation factor, $G^{2}(x_{\Pom},\bar{Q^{2}})$, in (\ref{eq:FLH}) 
compensates to a large extent the higher twist factor ${1\over Q^{2}}$ 
and the longitudinal SF remains large, and takes over $F_{T}^{D}$,
in a broad range of $Q^{2}$ of the practical interest, see Fig.~1.

In (\ref{eq:FLH}), the leading twist-4 term is the same as for NC diffractive DIS.
However, in the CC diffractive DIS, because non-conservation of weak 
current,  extra higher twist contributions to $F_{L}^{D(3)}$ come from 
the expansion of $B_{L}$ (always substantially dominated by the 
perturbative region). 

Further terms (both twist-4 and higher), come from 
the term $\propto A_{L}$ in (\ref{eq:DSigLT}). They 
receive large contributions from the low $k^2$ region.
In particular they assume a strong relevance for the charm--strange component
where terms $\propto m_c^2/Q^2$ appear. 
We find for the $A_L$ contribution: 

\beq
F^{D(4)}_L [A_L] \approx \frac{4 \pi}{9 \; \sigma^{pp}_{tot}} \;
\frac{(m_c^2 + m_s^2)}{m_c^2} \; 
\frac{\beta (1-\beta)^2}{Q^2 (1+\delta)} (1+2\beta+3\beta^2)
\left[ \alpha_S(\bar{Q}^2_L) G(x_{\Pom}, \bar{Q}^2_L) \right]^2
\label{eq:FLAL}
\eeq

Whereas the components coming from $A_L$ is low scale dominated, 
it gives comparable contributions as the leading twist-4 in the 
small $Q^2$ region. The overall dependence of $F_L^{D(3)}(cs)$ on 
$Q^2$ and its decomposition in the $A_L$ and $B_L$ components are 
shown in Fig.~2.

Due to the symmetry $z, (1-z)$ of Eqs.(\ref{eq:M2}, \ref{eq:AT} - \ref{eq:BL}), 
one finds similar results subject to the replacement 
$m_{c} \rightarrow m_{s}$ for the forward production of charm 
($F^D_{s}$) at $1-z \lsim m_{s}^{2}/m_{c}^{2}$.

\begin{figure}[htb] \label{FIGUR2}
\mbox{\epsfig{file=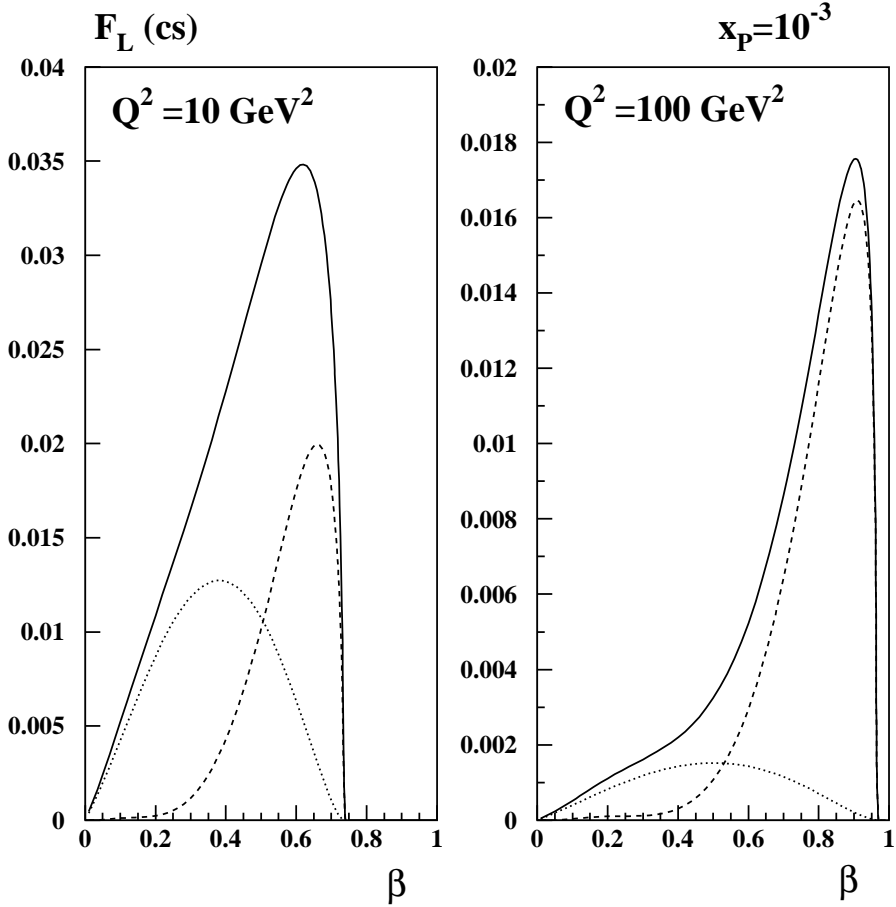,width=0.8\textwidth}}
\caption{$F_L^{D(3)}(cs)$ (solid) and $A_L$ (Eq.\ref{eq:FLAL}, dotted) and $B_L$ 
(Eq.\ref{eq:FLH}, dot--dashed) components of $ F_L$  at $Q^2$=10,100
GeV$^2$.
} 
\end{figure}

A relevant point is that the pQCD scales  $\bar{Q}^2{(c)}$ and
$\bar{Q}{(s)}^{2}$ are different, both explicitly depend on $\beta$, 
and the $x_{\Pom}$ and $\beta$ dependences of $F_{T(i)}^{D(3)}$ are 
inextricably entangled. This gives another example where
he Ingelman-Schlein factorization hypothesis, 
$F_{2}^{D(3)}(x_{\Pom},\beta,Q^{2}) = 
f_{\Pom}(x_{\Pom}) F_{2 \Pom}(\beta,Q^{2})$, with process
independent flux of pomerons in the proton $f_{\Pom}(x_{\Pom})$ and the 
$x_{\Pom}$ independent pomeron SF $F_{2\Pom}(\beta,Q^{2})$, is not confirmed 
by pQCD calculation. The diffractive factorization breaking in CC diffractive 
DIS is especially severe, because for the same $c\bar{s}$ final state the pQCD 
factorization scale $\bar{Q}^{2}$ changes substantially from the forward to 
backward hemisphere: $\bar{Q}_{s}^{2} \ll \bar{Q}_{c}^{2}$. Although the 
perturbative intrinsic charm component $F_{T(\bar{c})}^{D(3)}$ is suppressed by
the mass of a heavy quark, it is still substantial and it is predicted to rise 
much steeper than the strange one as $x_{\Pom}\rightarrow 0$. Furthermore, 
$F_{T(\bar{c})}^{D(3)}$ is truly of perturbative origin at all $\beta$, while 
$F_{T(s)}^{D(3)}$ has a non-negligible dependence to small scales up to 
$\beta \gsim 0.7$.

As an illustration of the diffractive factorization breaking, in
Fig.~3  we show the effective exponent of the $x_{\Pom}$ dependence
\beq
n_{eff} = 1-{\partial \log F_{2}^{D(3)} \over \partial\log x_{\Pom}}
\label{eq:neff}
\eeq
evaluated for $x_{\Pom}=3\cdot 10^{-3}$. 

\begin{figure}[htb] \label{FIGUR3}
\mbox{\epsfig{file=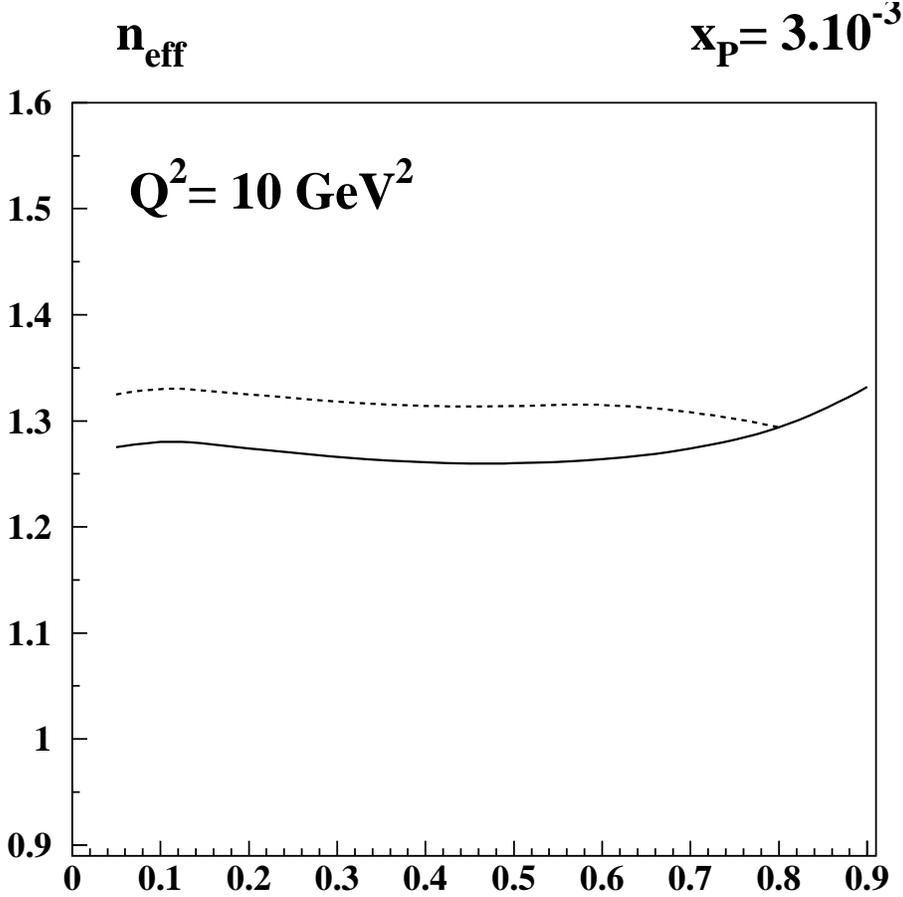,width=0.8\textwidth}}
\caption{
The $\beta$ dependence of $ n_eff$  for $x_{\Pom}=0.003$ and 
$Q^2=10$GeV$^2$. Solid line: ud component, dashed line: ud+cs  }
\end{figure}

Evidently, at fixed $\beta$, the $c\bar{s}$ excitation is possible
only for sufficiently large $Q^{2}$ such that $\beta_{c} > \beta$.
For this reason, diffractive SF's exhibit strong
threshold effects shown in Fig.~4, which are much stronger than
in the NC case studied in \cite{GNZcharm,HT4}. Notice, that $F_{3}^{D(3)}$
vanishes below the $c\bar{s}$ threshold.

\begin{figure}[htb] \label{FIGUR4}
\mbox{\epsfig{file=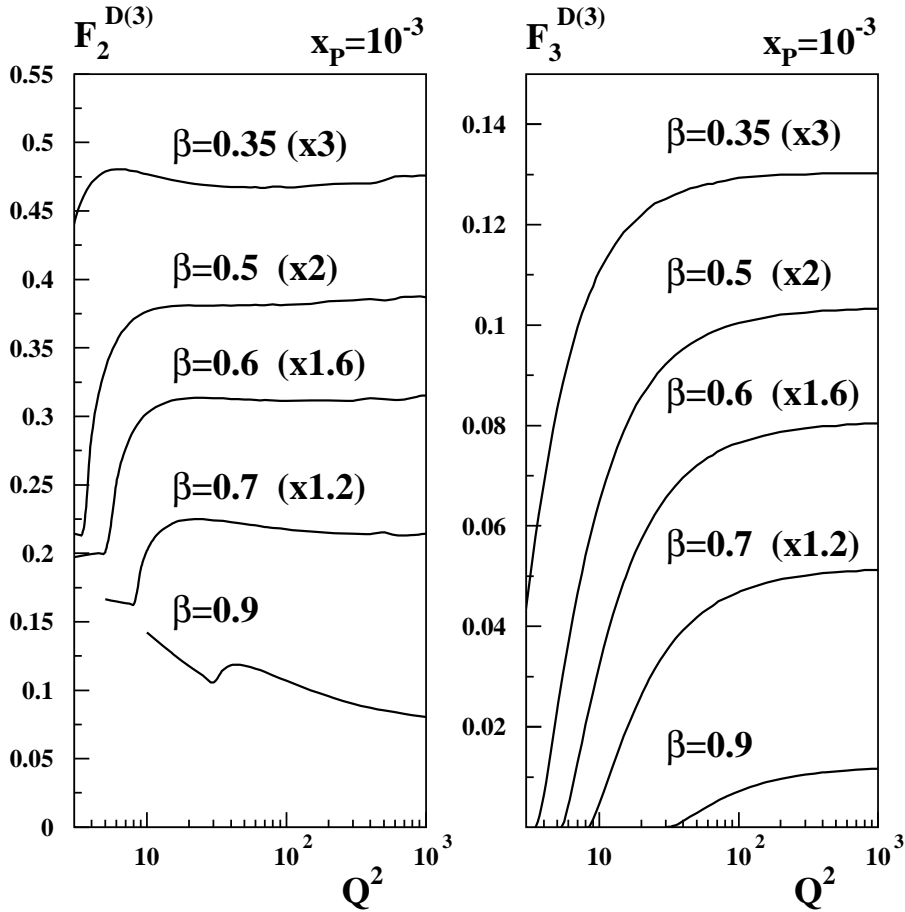,width=0.8\textwidth}}
\caption{Charm-strange threshold effect on Diffractive SF.
First box: $F_2^{D(3)}$. Second box: $F_3^{D(3)}$.
} 
\end{figure}

We have also calculated the contribution to
diffractive SF's from the top-(anti)bottom excitation.
With $m_{t}=180$ GeV and $m_{b}=4.5$ GeV, our result is that,
even for the very large $Q^2 = 10^4$ GeV$^2$, 
the $t\bar{b}$ contribution is about 50 times
smaller than the $c\bar{s}$ one, making very difficult
to observe this component at HERA.

Naively, one would expect $F_{3}^{D(3)}=0$ for a quark-antiquark
symmetric target  as the Pomeron is. Indeed, because $A_{3}(z)$, and
for equal mass case, $B_{3}(z)$ too, are antisymmetric
about $z={1\over 2}$, the contribution from
$u\bar{d}$ excitation to $F_{3}^{D(3)}$  vanishes upon the integration
over the $u$-jet production angles.
In contrast to that, in the $c\bar{s}$
excitation there is a strong forward-backward asymmetry, and
$F_{3}^{D(3)}=F_{T(s)}^{D(3)}-F_{T(\bar{c})}^{D(3)}\neq 0$.
Our predictions for $F_{3}^{D(3)}$ are shown in Fig.s~1 and 4.

Up to now we have discussed the intermediate and large $\beta$ regions,
the small $\beta$ one is, on the other hand, dominated by
the so-called triple-pomeron component.
In this case, diffraction proceeds via excitation
of the soft gluon-containing $q\bar{q}g$ and higher Fock states of
the photon. As it has been discussed to great detail in \cite{NZ94,GNZ95},
at $\beta \ll 1$ and only at $\beta \ll 1$, and with certain reservations,
one can apply the standard parton model treatment to diffractive DIS.
For instance, the conventional fusion of virtual photons with the gluon
from the two-gluon valence state of the pomeron becomes the driving term
of diffractive DIS.

For this component, the results for the diffractive SF of light
quarks coincide (once the opportune couplings of weak interaction are
substituted to the EM ones) with those presented in Ref. \cite{GNZ95}.
However, for the charm--strange part, it must be considered that
now a charm quark  always appears together with a strange one,
leading to a threshold ($Q^2_{cs}= 4 GeV^2$), which is intermediate
between the strange ($Q^2_{ss}= 1 GeV^2 $ ) and the charm
($Q^2_{cc}=10 GeV^2$) electromagnetic DIS thresholds in analogy to our
discussion concerning the usual DIS \cite{GNZcharm} (Referring 
to the notation of \cite{GNZ95} one finds $A_{cs}=0.08$.)

\bigskip


We have presented the calculation of charged
current diffractive structure functions
in QCD and carried on a comparison with electromagnetic case. 
Both charged current and electromagnetic
diffraction share the property of diffractive factorization breaking.
For instance, we find different $x_{\Pom}$ dependences of the intrinsic
$u,d$, strangeness and charm composition of the pomeron. Futhermore, we 
predict, for CC, even a 
different $x_{\Pom}$ dependence for the production of charm quark in the 
forward and backward direction.

Other new features of CC diffraction respect to EM one,
are the emergence of
substantial $F_{3}^{D(3)}$,  and the large higher twist contributions
to the longitudinal structure function. These predictions 
will be testable with the accumulation
of a larger statistic  on CC diffraction at HERA.



\begin{thebibliography}{299}
\bibitem{NZ92} 
N.N. Nikolaev  and B.G.~Zakharov, {\it Z. Phys.}
{\bf C53},  331 (1992).

\bibitem{NZsplit} 
N.N. Nikolaev and B.G.Zakharov,
{\sl Phys. Lett.} {\bf B332}, 177 (1994).

\bibitem{NZ94} 
N.N. Nikolaev  and B.G.Zakharov,  {\sl JETP}
{\bf 78}, 598 (1994);
{\sl Z. Phys.} {\bf C64} (1994) 631.

\bibitem{GNZ95} 
M. Genovese, N.N. Nikolaev  and B.G. Zakharov,
 {\sl JETP} {\bf 81} 625 (1995).

\bibitem{GNZA3Pom} 
M. Genovese, N.N. Nikolaev and B.G. Zakharov,
 {\sl JETP} {\bf 81},  633 (1995).

\bibitem{GNZcharm} 
M. Genovese M., N.N. Nikolaev  and B.G. Zakharov,
{\sl Phys. Lett.} {\bf B378}, 347 (1996).

\bibitem{GNZlong} 
M. Genovese, N.N. Nikolaev and B.G. Zakharov, {\sl Phys.Lett.}
{\bf B380},  213 (1996).

\bibitem{HT4}
M. Bertini, M. Genovese, N.N. Nikolaev, A.V. Pronyaev and B.G. Zakharov, 
{\it Twist-4 effects and $Q^2$ dependence of diffractive DIS}, 
hep-ph 9710547, to appear in {\sl Phys. Lett.} {\bf B}.

\bibitem{Bartels} 
J. Bartels, H. Lotter and M. W\"usthoff,
\PLB 379 (1996) 239.
H. Lotter, \PLB 406 (1997) 171.

\bibitem{Levin}
E.M.Levin, A.D.Martin, M.G.Ryskin and T.Teubner, {\it Z. Phys.}
{\bf C74} (1997) 671. E.Gotsman, E.Levin and U.Maor, {\it Nucl. Phys.}
{\bf B493} (1997) 354.

\bibitem{NZDIS97}
N.N. Nikolaev and B.G. Zakharov, DIS'96: Deep Inelastic Scattering and Related
Phenomena, Editors G.D'Agostini and A.Nigro, World Scientific,
Singapore, pp.347-353; Nikolaev N.N. and B.G.Zakharov, Phenomenology
of Diffractive DIS. Overview at DIS'97, Chicago, April hep-ph/9706343.

\bibitem{CCZEUS}
ZEUS Collab., M. Derrick et al., {Z. Phys.} {C72} (1996) 47.

\bibitem{CCDDexp}
J. Pliszka and A. F. Zarnecki, in "Future Phys. at
HERA", Hamburg 1996, pag. 728 ed. G. Ingelman {\it et al.}

\bibitem{ZEUSF2Pom} 
ZEUS: M. Derrick et al. {\sl Z. Phys.} {\bf C68}, 569 (1995)

\bibitem{F2charm}
V. Barone, M. Genovese, N.N. Nikolaev, E. Predazzi and B.G. Zakharov,
\ZPC 70 (1996) 83; Phys. Lett. {B304} (1993) 176, {B268} (1991) 279,
{B317} (1993) 433 ; V. Barone and M. Genovese, \PLB 379 (1996) 233.

\bibitem{F2cs}V. Barone, M. Genovese,
N.N. Nikolaev, E. Predazzi and B.G. Zakharov, \PLB 328 (1994) 143.

\bibitem{Buchmuller}
W. Buchm\"uller, M.F. McDermott and A. Hebecker, hep-ph/9703314.

\bibitem{NPZslope}
N.N. Nikolaev, A. Pronyaev and B.G. Zakharov, paper in preparation.

\bibitem{BGNPZcc} 
V. Barone et al., {\sl Phys. Lett.} {\bf B292}, 181 (1992).
V. Barone, M. Genovese, N.N. Nikolaev, E. Predazzi and B.G. Zakharov,
{\sl Int. J. Mod. Phys. } {\bf A8} (1993) 2779.

\bibitem{GRVNLO} 
M. Gl\"uck, E. Reya and A. Vogt, {\sl Z. Phys.} {\bf C67}, 433 (1995).


\end{thebibliography}
\end{document}